\documentstyle[11pt,newpasp,twoside,epsf]{article}
\def\edcomment#1{\iffalse\marginpar{\raggedright\sl#1\/}\else\relax\fi}
\reversemarginpar

\begin{document}
\title{Winds and Outflows in Starburst Galaxies and AGN}
 \author{Stefanie Komossa, G\"unther Hasinger}
\affil{Max-Planck-Institut f\"ur extraterrestrische Physik, Giessenbachstr.,
D-85748 Garching, Germany}
\author{Hartmut Schulz$^{\dagger}$}
\affil{Ruhr-Universit\"at Bochum, Astronomisches Institut, 44780 Bochum \newline
        $^{\dagger}$passed away in August 2003}

\begin{abstract}
Winds and outflows in starburst galaxies and AGN
provide important information on the the physics
of the `central engine', the presence and evolution
of (nuclear) starbursts, and the metal enrichment of the 
nuclear environment and the intergalactic medium.  
Here, we concentrate on two examples, X-ray observations of 
the (U)LIRG NGC\,6240 and the BAL quasar APM\,08279+5255.  
\end{abstract}

\section{Introduction}

We present recent results on winds and outflows in
starburst galaxies and active galactic nuclei (AGN) 
based on observations carried out with
the {\sl Chandra} X-ray observatory.

Outflows play a crucial role in 
enriching the nuclear environment with matter from the core
region, carrying metals and possibly dust. 
They thus represent an important component when
studying the recycling of interstellar and intergalactic matter.  

In starburst galaxies, superwinds are
driven by nuclear starbursts. Their study provides
important information on the evolution of these winds,
the metal content they carry, and the process of IGM
enrichment with metals and/or dust.

In AGN, outflows may be driven by radiation pressure
of the central continuum source.
Components in outflow include possible accretion-disk winds 
and ionized absorbers. 
{\sl Chandra} and {\sl XMM-Newton} spectra of AGN
with warm absorbers are rich in absorption features
which allow to study the ionized material,
its origin, evolution and interaction with the environment,
in great detail
(e.g., Simkin \& 
Roychowdhury 2002). 

\section{Starburst-driven Superwinds, AGN environment: NGC 6240}

\subsection{The galaxy NGC 6240}

Ultraluminous infrared galaxies (ULIRGs) are characterized by
their huge luminosity output in the infrared,
predominantly powered by super-starbursts and/or hidden AGN
(e.g., Genzel et al. 1998, Sanders et al. 1988, 1999).
Many distant {\sl SCUBA} sources, massive and dusty galaxies,
are believed to be ULIRG equivalents at high redshift.
{\em Local} ULIRGs
are therefore ideally suited to study the physics of galaxy formation
and evolution 
(many ULIRGs are mergers),
the processes of IGM enrichment, the physics of
superwinds
driven by the nuclear starbursts, and to asses the frequency
of hidden  
AGN at their cores.
The questions regarding the onset of starburst and AGN activity and their
evolution in mergers are of fundamental importance for our understanding
of AGN/black hole formation and evolution in general.

Here, we would like to concentrate on the
galaxy NGC\,6240, one of the nearest
members
of the class of (U)LIRGs, considered to be a key representative
of its class.
The galaxy is a merger in the process of forming
an elliptical galaxy.

{\sl ROSAT} observations (Schulz et al. 1998)
showed that NGC\,6240 is special in its
highly luminous extended X-ray emission, making it 
 one of the most luminous
X-ray emitters in extended emission known among field galaxies
(Komossa et al. 1998).

\subsection{{\itshape{Chandra}} results}

Given the complex nature of the X-ray emission of NGC\,6240 with
contributions from many components suggested from previous
X-ray observations, spatially resolved spectroscopy is crucial
to disentangle
all contributing components, determine their nature, and derive their
physical properties.
Here, we report results from the first spatially resolved X-ray
spectroscopy of NGC\,6240, carried out with the ACIS-S instrument
onboard the {\sl Chandra} X-ray observatory (see also Komossa et al. 2003).

The {\sl Chandra} image of NGC\,6240 reveals a wealth of structure,
changing in dependence of energy.
A large part of the X-ray emission of NGC\,6240 is extended, confirming previous
results first obtained with the {\sl ROSAT} instruments. 

Basically, the {\em extended} emission of NGC\,6240 shows three components:
firstly, some weak, widely extended halo emission, which was already hinted for
by our previous {\sl ROSAT} observations, not discussed further here. 
Secondly, an emission component spatially coincident with the H$\alpha$
emission of NGC\,6240. Thirdly, a more compact, harder, extended
component, likely related to some nuclear starburst activity.

In the {\sl Chandra} ACIS-S image, 
several X-ray `loops' and knots are visible which correlate
well with the H$\alpha$ emission of NGC\,6240 (Fig. 2), and 
which are most prominent between 1 and 2.5 keV.
Above 1.5 keV, X-ray emission from
the direction of the northern nucleus of NGC\,6240 starts to emerge.
The hard X-ray image is dominated by emission
from two compact sources, spatially coincident within the errors
with the IR positions of the two nuclei of NGC\,6240.

The extended X-ray emission of NGC\,6240 of
"butterfly-like" shape and coincident with the H$\alpha$ emission, 
is generally well described
by a MEKAL model with $kT \simeq 0.8$ keV and absorption with
a column density $N_{\rm H} \simeq 3\,10^{21}$ cm$^{-2}$.
Abundances are not well constrained, but fits typically yield
$Z \approx 0.1$\,solar when applying simple one-component 
thermal models.   

In addition to the widely extended emission, a more compact 
nuclear component, still extended, is present.
Representative of this emission, we have determined
the spectrum of an emission-blob south-west of the Southern nucleus
of NGC\,6240. It is well fit by a MEKAL model of temperature 
$kT \simeq 3 keV$ and abundances of $Z \approx$ solar.  

\begin{figure}[h]
\plotone{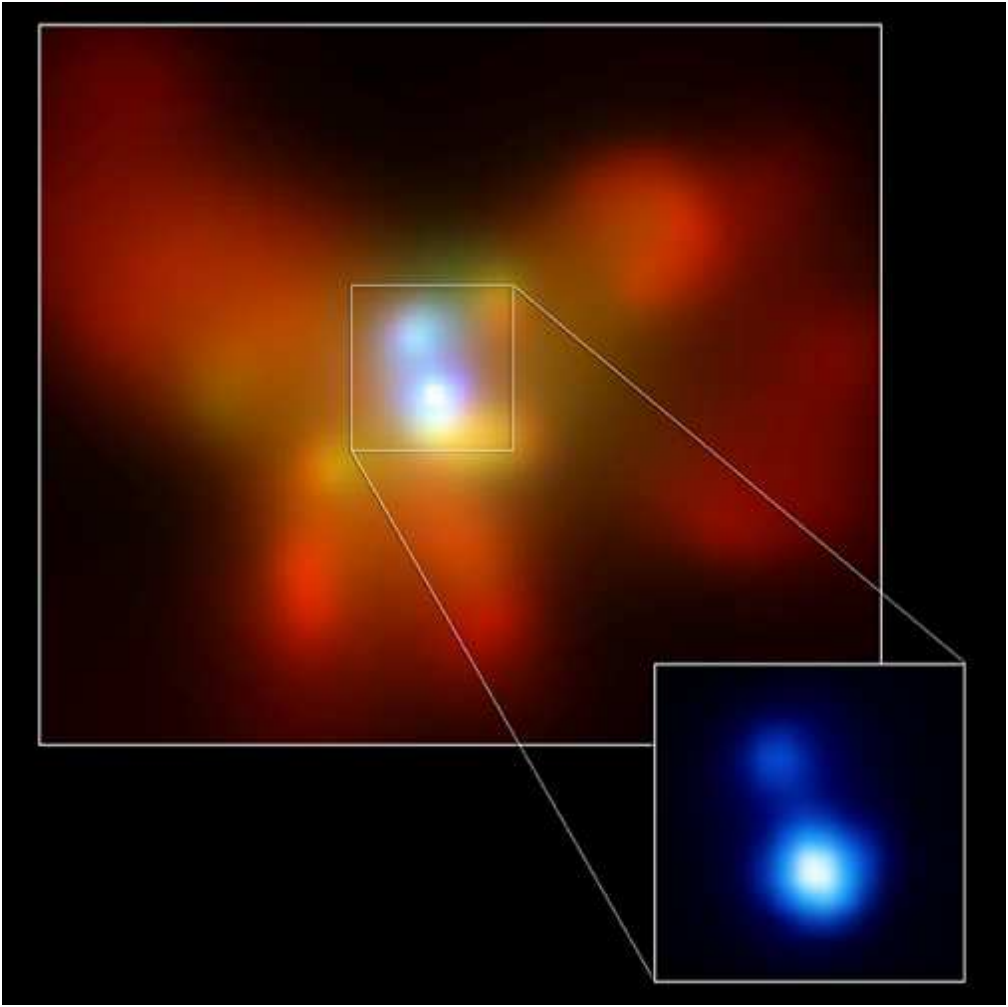}
\caption{X-ray energy image of NGC\,6240. The inset zooms onto the two hard nuclei. 
}
\end{figure}

The trend of decreasing abundances with increasing nuclear
separation of the starburst component of Mrk\,273 was noted
by Xia et al. (2002). 
Whether this trend is true for NGC\,6240, or whether the lower abundances 
inferred for the more extended starburst component of NGC\,6240 are still
an artifact of not applying more complex emission models (e.g., Komossa et al. 2000), 
has yet to be determined.
It would require much deeper {\sl Chandra} observation of NGC\,6240 
(and other (U)LIRGs) than presently
available, in combination with more complex, possibly non-equilibrium
emission models. 

According to superwind models of Schulz et al. (1998),
a mechanical input power of $L_{\rm mech} = 3\,10^{43}$ erg/s
(derived from a SN-rate of 3/year) can, within 3\,$10^7$ yrs,
drive a single shell to an extent of $R \sim$ 10\,kpc
within a medium of 0.1 cm$^{-3}$ particle density.
This model reproduces the observed X-ray luminosity
of the extended, butter-fly shaped,
starburst-related X-ray emission of NGC\,6240 of $\sim$10$^{42}$ erg/s.

The spectra of both X-ray nuclei of NGC\,6240 are very hard in
X-rays and show strong, $\sim$neutral iron K\,$\alpha$ lines.
These properties
identify both nuclei of NGC\,6240 as active, and
suggest a scattering geometry for {\em both} AGN in NGC\,6240
(the core components are discussed in greater detail by Komossa et al. 2003).
In addition to the 6.4 keV iron line, which is traced back
to fluorescence from cold material in the
AGN environment, a weaker line from highly ionized
iron shows up, which could be related to supernova remnants or
an ionized scatterer (see Komossa et al. 1998 for models
of a warm scatterer in NGC 6240, and Boller et al. 2003 for
recent {\sl XMM} data which resolved the high-ionization iron-line
complex in two components.)


\begin{figure}[h]
\plotone{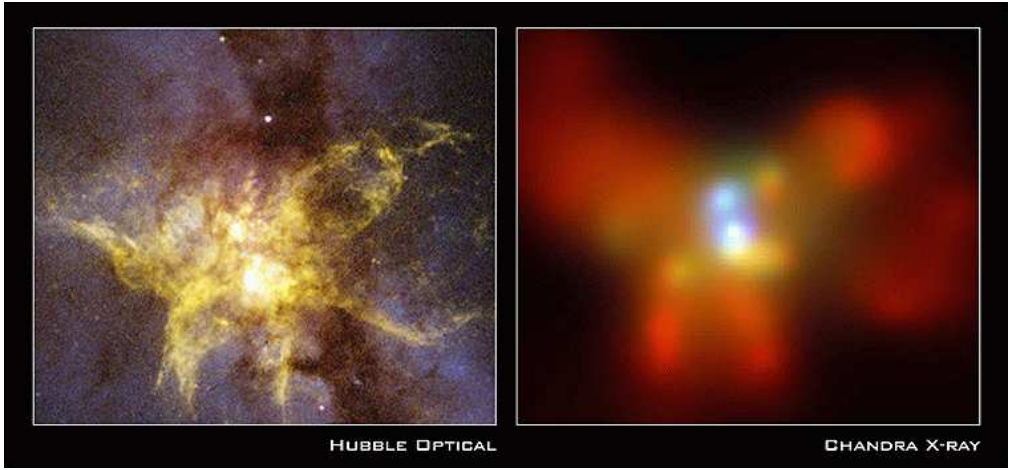}
\caption{Comparison of the H$\alpha$ image (Gerssen et al. 2001) with the 
  X-ray image of NGC\,6240 [NASA/CXC/Komossa et al. 2003].}
\end{figure}

\section{Absorption and outflows in AGN}

Cold and `warm' absorption is ubiquitous in the AGN environment.
X-ray absorption and emission features provide
valuable diagnostics of the physical conditions
in the X-ray gas and, in particular, allow to measure elemental
abundances at high redshift, with profound consequences for
our understanding of the star formation history in the early universe.

\subsection{BAL quasars}
Here, we would like to concentrate on BAL quasars. 
They are characterized by broad UV absorption lines.
It has been suggested that these lines arise in
a flow of gas which rises vertically from
a narrow range of radii from the accretion disk. The flow then bends
and forms a conical wind moving radially outwards (Elvis 2000).
Variants of radiatively-driven disk-winds were explored
(e.g., Proga et al. 2000, and references therein).
In some of these models, an X-ray absorber shields the wind downstream
from soft X-rays, allowing resonant-line driving to remain effective and
accelerate the outflowing BAL wind up to $\sim$0.1c. 

These high-velocity BAL winds may play a crucial role in metal-enrichment
of the interstellar medium of the galaxy, possibly even the IGM.  

\subsection{The quasar APM\,08279+5255}

The high-redshift BAL quasar APM 08279+5255 is magnified by a gravitational lense and
is among the most luminous objects in the universe (even after
correction for lensing). It  
has one of the best-measured X-ray spectra among BALs.
The recent {\sl XMM-Newton} observation led to the detection
of a strong absorption feature of ionized iron, interpreted as K-edge,
arising from a warm
absorber of high column density (Hasinger et al. 2002),
and variable on short timescales, when compared with the {\sl Chandra}
observation (Chartas et al. 2002). 

We concentrate here on the aspect of deriving metal abundances
at high redshift (for an application to measure cosmological
parameters, see e.g. Alcaniz et al. 2003).
Using our best-fit warm absorber model of the {\sl XMM} spectrum
of APM 08279+5255, we infer an abundance of Fe/O $\simeq$ 3 $\times$ solar   
which is rather high, given the high redshift of
the object.

The element iron plays a special role in
chemical enrichment scenarios, because its production
is delayed relative to other elements (e.g., Fig 1 of Hamann
\& Ferland 1993), since it is believed to be mostly produced
in supernovae of type Ia. 
Taken at face value, the {\sl XMM} observation
of APM 08279+5255 imply an efficient iron
production mechanism in the early universe. 
The unusual strength of the iron features and the indications
for strong variability make APM 08279+5255 an excellent target
for simultaneous, deep follow-up observations with {\sl Chandra}
and {\sl XMM-Newton}.

While Fe X-ray features have only been observed in very few BALs so
far, with future X-ray missions like {\sl XEUS} (Arnaud et al. 1998)
we will be able to observe more objects
with high accuracy.
The iron absorption lines and K-edges provide a unique
probe of matter at high redshift because, firstly, they are easy
to measure even, or, {\em particularly}, at high redshift $z$,
and secondly, Fe(/O) abundance measurements in the early
universe are important for constraints on the early star formation history. 

Science issues that we will be able to address in detail with
future X-ray missions for the first time,  
particularly at high redshift, include 
the determination of metal abundances
of X-ray absorbers by
detection of metal absorption edges,
the determination of the composition of dust mixed with cold and ionized
gas (K-edges of metals in cold dust and cold gas will be resolvable from each other for the first time),
measurements of the velocity field of the gas,
and the utilization of these results to investigate the effects of reprocessing
of gas and dust
in AGN from high to low redshift. 

\acknowledgments{I, St.K., am grateful to  Hartmut Schulz
for intense collaborations on NGC 6240 ever since I started in astrophysics,
for sharing his wide knowledge, and for 
ongoing discussions on many topics of astrophysics. 
Hartmut passed away in August 2003, after a short, severe illness. \newline
We thank R. van der Marel for providing the HST H$\alpha$ image of NGC\,6240. }


\begin{references}
\reference Alcaniz J.S., Lima, J.A.S., Cunha, J.V. 2003, MNRAS, 340, L39 
\reference Arnaud, M., et al. 1999, {\sl The XEUS Science Case}, ESA
\reference Boller, T., et al. 2003, astro-ph/0307326
\reference Chartas, G., Brandt, W.N., 
                 Gallagher, S.C., Garmire, G.P., 2002, \apj, 597, 169    
\reference Elvis, M., 2000, \apj, 545, 63 
\reference Genzel, R., Lutz, D., Sturm, E., et al. 1998, \apj, 498, 579
\reference Gerssen, J., van der Marel, A.P., Axon, D., Mihos, C.,
           Hernquist, L., Barnes, J.E., 2001, ASP Conf. Ser., 249, 665 
\reference Hamann, F., Ferland, G.J. 1993, \apj, 418, 11 
\reference Hasinger, G., Schartel, N., Komossa, S. 2002, \apj, 573, L80 
\reference Komossa, S., Schulz, H., Greiner, J. 1998, \aap, 334, 110  
\reference Komossa, S., Breitschwerdt, D., B\"ohringer, H., Meerschweinchen, J. 2000, Ap\&SS, 272, 295 
\reference Komossa, S., Burwitz, V., Hasinger, G., Predehl, P., 
           Kaastra, J.S., Ikebe, Y., 2003, \apj, 582, L15 
\reference Proga, D., Stone, J.M., Kallman, T.R., 2000, ApJ, 543, 686
\reference Sanders, D.B., Soifer, B.T., Elias, J.H., et al. 1988, \apj, 325, 74   
\reference Sanders, D.B. 1999, Ap\&SS, 266, 331
\reference Schulz, H., Komossa, S., Bergh\"ofer T., Boer, B. 1998, \aap, 330, 823 
\reference Simkin, M.V., Roychowdhury, V.P., 2002, cond-mat/212043  
\reference Xia, X.Y., Xue, S.J., Mao, S., et al. 2002, \apj, 564, 196   
\end{references}
\end{document}